\documentclass[modern]{aastex62}

\usepackage{fontawesome}
\definecolor{linkcolor}{rgb}{0.1216,0.4667,0.7059}

\newcommand{\commentlink}[1]{\href{https://github.com/KeplerGO/ScientificOpportunities/issues/#1}{\sc \faExternalLink\ Discuss this topic}\,\,}

\graphicspath{{./}{figures/}}

\shorttitle{Kepler's discoveries will continue}
\shortauthors{Barentsen et al.}

\begin{document}

\title{\fontsize{14}{17} \sc
Kepler's discoveries will continue:\\
\fontsize{14}{20} 21 important scientific opportunities\\
\fontsize{14}{20} with Kepler \& K2 archive data
\vspace{5mm}
}

\correspondingauthor{Geert Barentsen}
\email{geert.barentsen@nasa.gov}

\author{Geert Barentsen}
\affiliation{Kepler/K2 GO Office, BAER, P.O. Box 25, Moffett Field, CA 94035, USA}
\affiliation{NASA Ames Research Center, MS 244-30, Moffett Field, CA 94035, USA}

\author{Christina Hedges}
\affiliation{Kepler/K2 GO Office, BAER, P.O. Box 25, Moffett Field, CA 94035, USA}
\affiliation{NASA Ames Research Center, MS 244-30, Moffett Field, CA 94035, USA}

\author{Nicholas Saunders}
\affiliation{Kepler/K2 GO Office, BAER, P.O. Box 25, Moffett Field, CA 94035, USA}
\affiliation{NASA Ames Research Center, MS 244-30, Moffett Field, CA 94035, USA}

\author{Ann Marie Cody}
\affiliation{Kepler/K2 GO Office, BAER, P.O. Box 25, Moffett Field, CA 94035, USA}
\affiliation{NASA Ames Research Center, MS 244-30, Moffett Field, CA 94035, USA}

\author{Michael Gully-Santiago}
\affiliation{Kepler/K2 GO Office, BAER, P.O. Box 25, Moffett Field, CA 94035, USA}
\affiliation{NASA Ames Research Center, MS 244-30, Moffett Field, CA 94035, USA}

\author{Steve Bryson}
\affiliation{NASA Ames Research Center, MS 244-30, Moffett Field, CA 94035, USA}

\author{Jessie L. Dotson}
\affiliation{NASA Ames Research Center, MS 244-30, Moffett Field, CA 94035, USA}

\begin{abstract}
NASA's Kepler Space Telescope has collected high-precision, high-cadence time series photometry on 781,590 unique postage-stamp targets across 21 different fields of view. These observations have already yielded 2,496 scientific publications by authors from 63 countries. The full data set is now public and available from NASA's data archives, enabling continued investigations and discoveries of exoplanets, oscillating stars, eclipsing binaries, stellar variability, star clusters, supernovae, galaxies, asteroids, and much more.

In this white paper, we discuss 21 important data analysis projects which are enabled by the archive data. The aim of this paper is to help new users understand where there may be important scientific gains left to be made in analyzing Kepler data, and to encourage the continued use of the archives. 
With the TESS mission about to start releasing data, the studies will inform new experiments, new surveys, and new analysis techniques. The Kepler mission has provided an unprecedented data set with a precision and duration that will not be rivaled for decades. The studies discussed in this paper show that many of Kepler's contributions still lie ahead of us, owing to the emergence of complementary new data sets like Gaia, novel data analysis methods, and advances in computing power. Kepler's unique data archive will provide new discoveries for years to come, touching upon key aspects of each of NASA's three big astrophysics questions; How does the universe work? How did we get here? Are we alone? 
\end{abstract}


\vspace{10mm}

\section{Introduction} \label{sec:intro}

NASA's Kepler Space Telescope revolutionized our understanding of the universe. Launched in 2009 to detect Earth-sized planets around other stars \citep{borucki2010}, Kepler's high-precision and high-cadence photometry provided a new way for astronomers to study the sky. Kepler established that planets are ubiquitous \citep[e.g.][]{borucki2011,batalha2013,burke2015}, discovered that the properties and configurations of those planets are diverse \citep[e.g.][]{lissauer2011}, revealed the properties and interior structure of stars \citep[e.g.][]{bedding2011,beck2012,chaplin2014}, elucidated extragalactic transients \citep[e.g.][]{garnavich2016,rest2018}, and enabled studies of a variety of new and unusual types of astrophysical phenomena \citep[e.g.][]{welsh2011,tabby2016}.

Between 2014 and 2018, Kepler expanded and diversified its data set by carrying out a second mission named K2 -- the ``two reaction wheel Kepler'' \citep{howell2014}.  During this extended phase of the mission, the spacecraft surveyed an additional 20 fields along the ecliptic plane. The K2 mission tripled the total number of targets observed by Kepler. It enabled new investigations by observing objects which were not the focus of the original mission, including
young disk stars \citep[e.g.][]{ansdell2016},
white dwarfs \citep[e.g.][]{hermes2017},
microlensing events \citep[e.g.][]{henderson2016},
and extragalactic transients \citep[e.g.][]{rest2018}. All these targets were chosen by peer review and, owing to Kepler's generous pixel budget, included high-risk, high-reward science programs \citep[e.g. the search for planets around white dwarfs;][]{sluijs2018}. 

\begin{figure}

\vspace{10mm}

\plotone{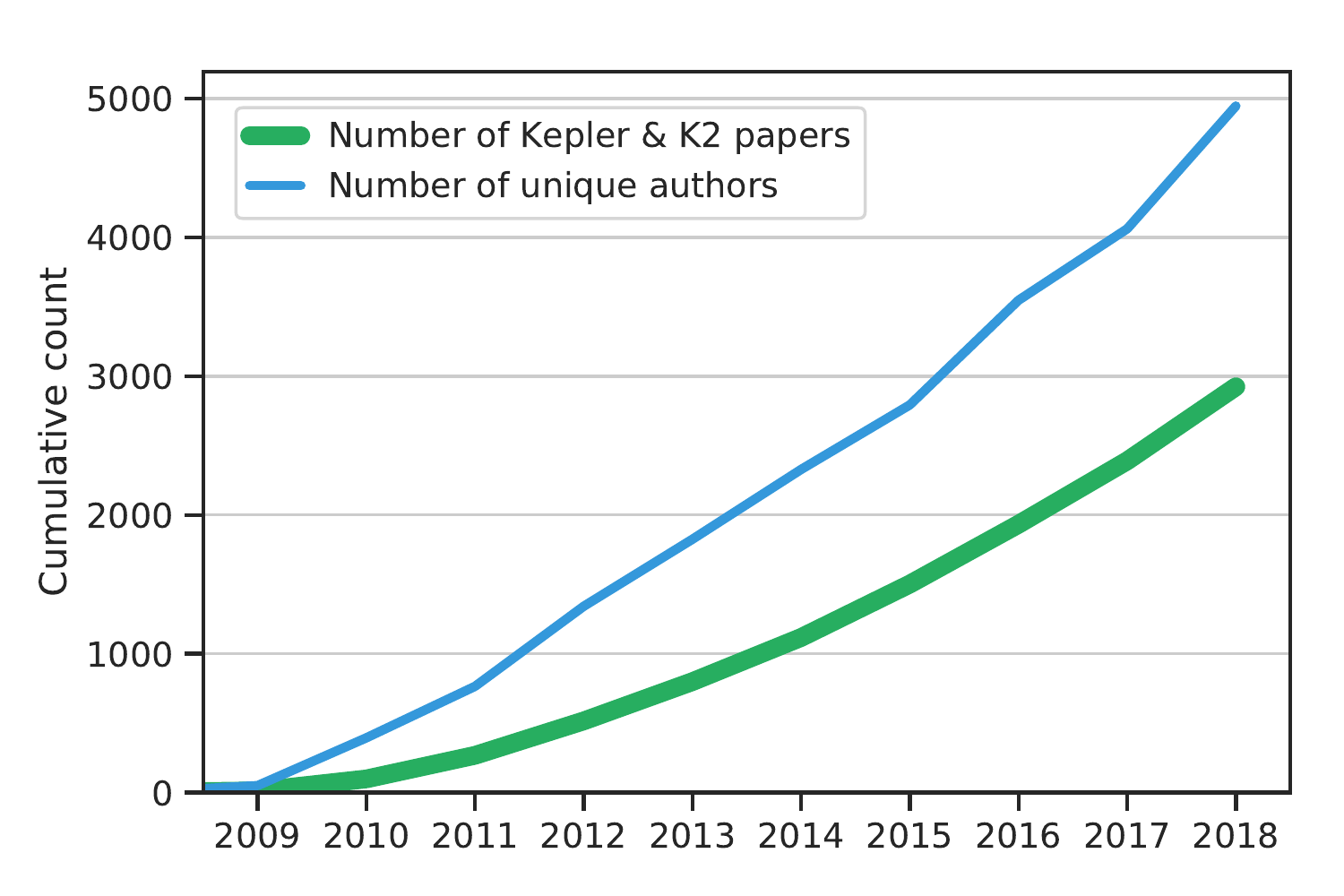}
\caption{Cumulative number of Kepler \& K2 publications and unique authors over time.  The full list of publications is available at \url{https://keplerscience.arc.nasa.gov/publications.html}.\label{fig:publications}}

\vspace{10mm}

\plotone{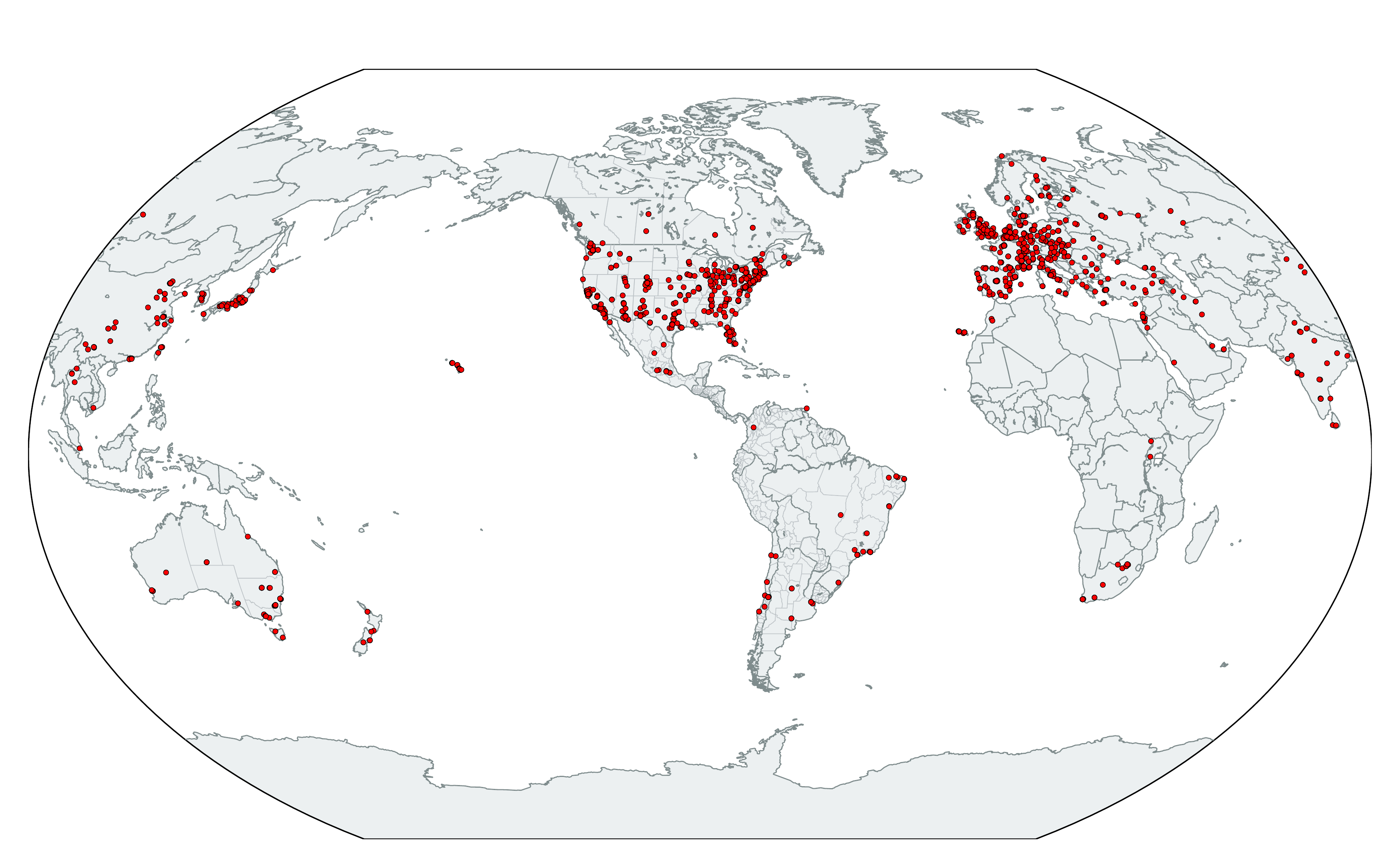}
\caption{Map of the institutions of authors and co-authors of Kepler and K2 publications. To date, Kepler data have been used in 63 countries across 6 continents.\label{fig:map}}
\end{figure}

To date, Kepler and K2 have contributed to 2,496 scientific publications authored by 4,968 unique authors across 63 countries and 6 continents (Figs.~\ref{fig:publications} \& \ref{fig:map}). These publications have been cited 81,912 times and include 58 PhD theses in the US alone. Kepler's scientific productivity has grown every year since launch, currently averaging 2 new papers per day. With 532 publications so far this year, 2018 is already the most successful year on record and brings Kepler on par with NASA's Great Observatories. 

The scientific impact of Kepler data has been enabled by the large dataset of 781,590 unique target masks\footnote{Kepler stored postage stamp images for pre-selected targets. There are a total of 781,590 unique post stamp identifiers for which data has been collected during one or more Kepler Quarter or K2 Campaign. A typical mask contains exactly one star, but $\sim10\%$ of the masks were scheduled to cover extended regions such as star clusters and galaxies. A legacy catalog detailing all the objects observed by Kepler is in preparation.} containing plethora stars and a sample of asteroids, star clusters, and galaxies. The dataset spans a range of stellar spectral types and ages. Even though much of the data collected during Kepler's original 4-year mission have been analyzed thoroughly by the mission pipeline \citep[e.g.][]{jenkins2010,twicken2016,thompson2018}, the scientific community continues to extract new discoveries from the archive data \citep[e.g.][]{shallue2018}. The K2 mission tripled the total number of targets observed, enabling continued scientific discoveries as new data have been released and new techniques developed.

New results from Kepler data have been enabled by novel algorithms \cite[e.g.][]{foreman-mackey2017,hedges2018,hon2018}, complementary new data sets including spectroscopic surveys and Gaia \citep[e.g][]{wittenmyer2018,zong2018,berger2018}, and the emergence of new scientific questions \citep[e.g.][]{teachey2018}.  Many recent results have benefited from an improved understanding of the best data analysis practices, advances in computing power, the emergence of collaborative hack days, and the availability of accessible open source software tools (e.g. \texttt{Astropy} and \texttt{Lightkurve}). Notably, K2's open data policy lowered the barrier for early-career researchers and citizen scientists to enter the growing field of exoplanet science. Over the past 4 years alone, the number of unique authors in Kepler's vibrant community doubled from $\sim$2,500 to $\sim$5,000 scientists (Fig.~\ref{fig:publications}).

In this paper, we argue that Kepler's discoveries are likely to continue for many years.  We present a non-exhaustive list of 21 data analysis projects that can be carried out using public Kepler and K2 data which are readily available in the data archives at MAST\footnote{\url{https://archive.stsci.edu}} or the NASA Exoplanet Archive\footnote{\url{https://exoplanetarchive.ipac.caltech.edu}}.  The aim of this paper is to help new users understand where scientific gains may be made, and to encourage the continued use of Kepler archive data.

We have made every effort to cite existing studies that were known to us at the time of publication, and invite community feedback\footnote{You can help edit the paper at \url{https://github.com/KeplerGO/ScientificOpportunities}} to help us credit additional works we have overlooked. This paper is not intended to be a comprehensive literature review of Kepler's 2,946 publications, and many prominent and impactful works by the Kepler science community have not been included in this paper as a result.\\

\section{scientific opportunities}

Kepler's community has been vibrant, innovative, and has produced science at a spectacular pace. 
All the projects listed in this paper have already been explored by talented teams. Certain aspects of these projects have remained unsolved -- not because of a lack of drive or expertise, but primarily for the following reasons:

\begin{enumerate}
\item K2's open data policy and its fast-paced delivery schedule (new data was released approximately every 3 months) have encouraged users to pursue quick, high-impact discoveries first. Projects which are more time-consuming have been lower priority, potentially meaning several more difficult discoveries and analyses are awaiting attention from the community.
\item Some projects could not be completed until the final data products had been released. For example, occurrence rate studies require complete, uniform, and carefully-characterized planet catalogs. Other studies are waiting for the K2 uniform global re-processing effort to complete\footnote{\url{https://keplerscience.arc.nasa.gov/k2-uniform-global-reprocessing-underway.html}}.
\item Some projects were not computationally tractable in the early days of Kepler. New data analysis techniques (such as easily accessible machine learning tools) and more computational power allows them to be executed now.
\item Some data products can be challenging to work with. Challenges include large file sizes and complicated data formats. The Kepler Guest Observer (GO) Office have recently begun releasing new High Level Science Products \citep[e.g. mosaics of star clusters,][]{cody2018} and new tools and tutorials (see the \texttt{Lightkurve}\footnote{\url{https://docs.lightkurve.org}} Python package) to enable the community to better access Kepler data and expertise, including the more unusual data products. 
\end{enumerate}

Below we summarize 21 science opportunities (11 exoplanet, 6 galactic, 2 extragalactic, and 2 solar system science projects). They have been inspired by a multitude of existing recent literature publications and conference presentations. While we have made every effort to cite these works, we remind the reader that this paper does not intend to be a comprehensive literature review.

The end of each section below contains a ``\textsc{Discuss this topic}'' button, which links to an issue on the GitHub repository of this paper where we invite researchers to discuss their ideas or progress towards resolving the challenge.\\

\subsection{Exoplanet science projects}

\subsubsection{Building a homogeneous catalog of K2 planets}
Homogeneous planet catalogs enable the accurate study of the occurrence rates of planets. Although the first four years of Kepler data have been searched thoroughly by the official Kepler pipeline \citep{jenkins2010}, providing a complete and well-characterized catalog \citep[][and references therein]{thompson2018}, the search for planets in the K2 phase of the mission was left to the community \citep[e.g.][]{foreman-mackey2015,montet2015,barros2016,crossfield2016,pope2016,dressing2017,vanderburg2016,luger2016,luger2018,petigura2018,livingston2018,mayo2018,yu2018}.

Planet searches in the K2 dataset by the community are driven primarily by planets that are high value for follow-up, either from the ground or from space. These high-value targets tend to be bright stars, with larger planets. This has caused the K2 exoplanet catalog to be less homogeneous and less complete than than the original Kepler mission catalog. A reliable, complete, and well-characterized catalog of K2 planets across all Campaigns would enable new planet occurrence rate studies, and to fully identify and rank the best planets for follow-up observations (e.g. atmosphere observations with the Hubble or James Webb Space Telescopes).  The K2 planet sample is expected to complement that of TESS, by adding smaller and cooler planets on longer orbits, owing to Kepler's higher precision.\\
\begin{center}
\commentlink{1}
\end{center}
\ \\

\subsubsection{Continuing work to identify planet occurrence rates as a function of stellar age, stellar type, and environment} \label{sec:occRate}

Accurately measuring occurrence rates of exoplanets not only provides insight into the prevalence of Earth-like planets in the universe, but also allows us to better design future missions for planet characterization. While the occurrence rates of planets have been studied carefully by several teams using data from the original Kepler field \citep[e.g.][and references therein]{burke2015,mulders2018,garrett2018}, planet occurrence rates are yet to be estimated in detail using the K2 data set.  Interestingly, K2 provided access to a wider range stellar ages \citep[e.g.][]{mann2017}, later stellar types \citep[e.g.][]{dressing2017}, and different Galactic populations.  The astrophysical diversity of the K2 data may reveal variability in the frequency of planets as a function of their environment. This, in turn, may inform planet formation models and future mission designs \citep[see][]{kopparapu2018}.

Moreover, while there have been numerous occurrence rate studies using the original Kepler data, its final Data Release 25 (DR25) planet catalog products have only recently become available and have thus only been utilized by a limited number of studies \citep{mulders2018,narang2018,petigura2018-2}. DR25 is the first Kepler planet catalog to be accompanied by an accurate characterization of the detection reliability and completeness \citep{coughlin2017,thompson2018}
and provides an important opportunity for improved occurrence rate studies.
The associated documentation recently became easier to access via the new  {\em Kepler Data Products Overview} page\footnote{\url{https://exoplanetarchive.ipac.caltech.edu/docs/Kepler_Data_Products_Overview.html}} at the NASA Exoplanet Archive.
\\
\begin{center}
\commentlink{2}
\end{center}
\ \\

\subsubsection{Discovering planets in background stars and under-utilized masks}
Many of the pixel masks observed by Kepler contain more than one star. These overlooked background stars potentially present opportunities to find new exoplanet candidates.  Most planet search pipelines focus on the primary target in the center of the mask, often ignoring neighbor stars towards the mask edges (exceptions include K2SC, \citealt{aigrain2016}).

Kepler's data archive also contains 4,160 extended pixel masks which were observed during Kepler Quarter 5 through 17 to estimate the background Eclipsing Binary rate\footnote{To access these masks, search for {\em Investigation ID = EXBA} in the Kepler data search interface at \url{https://archive.stsci.edu/kepler/data_search/search.php}.}. These masks have not been searched for planets by the Kepler pipeline. In addition, K2 collected a significant number of extended masks to observe star clusters, galaxies, and moving Solar System object.  All these masks contain a significant number of background stars which, to our modest knowledge, have not been analyzed or searched for planets thoroughly. Superstamps from Kepler and K2 can be difficult to work with, due to them being built from a set of smaller tiles. Recently, the Kepler GO office has started to release easier-to-use mosaics of these data \citep{cody2018} and tools to cut out small Target Pixel Files from such superstamps (cf. \texttt{Lightkurve}).
\\
\begin{center}
\commentlink{3}
\end{center}
\ \\

\subsubsection{Discovering planets in crowded regions}
Crowded fields, where flux from multiple stellar sources falls into the same pixel region, can be difficult to search for exoplanet signals. Flux from several sources dilutes the signal from the exoplanet transit, making them harder to detect. This is particularly hard using the Simple Aperture Photometry (SAP) method typically used in Kepler analyses.

Because Kepler's original mission strategically focused on isolated stars, the K2 and TESS communities are still developing the tools and expertise required to effectively extract science from blended stars in crowded regions. Point Spread Function (PSF) fitting photometry has already been shown to be a viable route towards extracting science from K2 \citep{libralato2016,libralatob2016,nardiello2016}, but the technique has only been applied to a subset of K2 cluster data so far.  In addition, difference imaging has been applied to the crowded K2 Campaign 9 field towards the Galactic Bulge \citep{wang2017} and towards the M35 cluster \citep{soares2017}. A comprehensive search for planets using these alternative photometry techniques may reveal new planets in such interesting regions.
\\
\begin{center}
\commentlink{4}
\end{center}
\ \\

\subsubsection{Discovering planets around binary stars}
Investigating the frequency of planets around binary stars was one of Kepler's prime mission goals, as binary stars and multi-star systems are common.  Although some planets have been found to orbit binary stars \cite[e.g. Kepler-16b,][]{doyle2011}, only 11 have been discovered so far \citep{fleming2018}.
The identification of planets in binary star systems is challenging because i) their complicated gravitational interactions may lead to irregular orbital periods and ii) the flux from two stars dilutes transits, reducing their depths. More work is needed to understand our biases and sensitivity of binary star planets and constrain their frequency.
We are not currently aware of publications discussing dedicated searches or occurrence rates for planets in multiple systems. This is nevertheless important because a large fraction of the stars in our Galaxy are thought to occur in multiple systems \citep{duchene2013}. 
 \\
\begin{center}
\commentlink{5}
\end{center}
\ \\

\subsubsection{Discovering planets with unusual transit shapes} 
The shape of an exoplanet transit provides critical information on the size and shape of the transiting object. Candidate objects that may exhibit different transit shapes, such as disintegrating planets, exocomets, dust clouds and planets with rings, have already been discovered in Kepler \citep[e.g.][]{vanderburg2015, tabby2016, rappaport2018} and in other surveys \citep[e.g.][]{mamajek2012}. These objects can greatly further our understanding of planet diversity and planet formation mechanisms. Due to most transit searches assuming a ``box-like'' transit shape, some transiting objects may have escaped detection due to an unusual transit shape. Such objects may be difficult to detect using standard planet search methods, and may benefit from dedicated searches.
\\
\begin{center}
\commentlink{6}
\end{center}
\ \\

\subsubsection{Discovering planets using forward modeling techniques} \label{sec:fwModelPlanets}
To find planet signals of small planets around solar-like stars, we require data analysis methods that can accurately remove instrument systematics. In recent years, new planet detection methods have appeared which leverage advances in computing power and forward modeling techniques to detect planet candidates by simultaneously modeling planet signatures and instrument systematics \citep[e.g.][]{foreman-mackey2015, luger2016, luger2018}.  These data-driven models isolate and remove instrumental signals, preserving light curves of transiting exoplanets. To date, comprehensive catalogs which utilize such innovative planet detection methods have not been published.
\\
\begin{center}
\commentlink{7}
\end{center}
\ \\

\subsubsection{Discovering planets using Campaign 9 microlensing data}
Microlensing surveys offer a way to detect planets at very large separations from their host stars, exploring an area of parameter space that is difficult to achieve with either the transit method or the radial velocity method \citep[e.g.][]{penny2018}. During K2 Campaign 9, Kepler monitored an area of 3.7 square degrees towards the Galactic Bulge \citep{henderson2016}.  During this ~70-day Campaign, several dozen exoplanet microlensing events are known\footnote{A list is available via the NASA Exoplanet Archive's ExoFOP system: \url{https://exofop.ipac.caltech.edu/k2/microlensing/}} to have been observed simultaneously from space and from the ground \citep[e.g.][]{zhu2017,kim2018}. The resulting parallax measurements should in principle allow for the direct measurement of the masses of and distances to the lensing systems, thereby resolving degeneracies. To date this dataset has not been used to detect transiting planets. The K2 Campaign 9 data is challenging in nature owing to the data formats, the motion systematics, and the crowding \citep[e.g. see][]{poleski2018}. We also note that the C9 data set can be used to reveal variables of all types towards the Galactic Bulge, including distance markers such as RR Lyrae stars.
\\
\begin{center}
\commentlink{8}
\end{center}
\ \\

\subsubsection{Identifying Transit Timing Variations in overlapping K2 fields}
Transit Timing Variations (TTVs) can identify multi-planet systems, identify dynamically interesting systems and provide insight on the masses of planets.  Existing planet search algorithms, which are designed to search for periodic signals, may not be sensitive to planets which show extreme TTVs. Identification and analyses of TTVs in K2 are enabled by the various overlapping fields. In particular the K2 Legacy Field (where K2 visited the same field in Campaigns 5, 16, and 18) provides new scope to search for TTVs. Additionally, in the future there will be the opportunity to increase this baseline when TESS overlaps with Kepler fields \citep[e.g. see][]{barclay2018}. The forecasting of interesting TTV's that can be found using Kepler and TESS combined has already begun \citep[e.g.][]{goldberg2018, callista2018}.
\\
\begin{center}
\commentlink{9}
\end{center}
\ \\

\subsubsection{Discovering planets on year-long orbits by mitigating local background variations} \label{sec:oneYearOrbits}
Discovering small planets on year-long orbits informs our understanding of our place in the universe by helping to establish how common Earth-like planets may be in the universe. Kepler's ability to discover these planets has been limited by the presence of systematics with similar year-long periodicities. Certain CCD channels (in particular channels 26, 44 and 58) experience the “rolling band” effect, where the background shows a strong time-varying component appearing as bands moving across the detector \citep[see \S6.7 of][]{vancleve2016}.

The rolling band artifact often adds spurious signals which mimic small planet transits. Because temperature variations trigger the artifact, the presence and characteristics of the rolling bands are correlated with Kepler's own $\sim$year-long orbit around the Sun, leading to a significant excess of false positive planet candidates on $\sim$year-long orbits \citep{thompson2018}.

Local background estimation techniques may help to remove the systematic\footnote{e.g. see \url{https://docs.lightkurve.org/tutorials/2.06-identify-rolling-band.html}}, but the Kepler pipeline only applied global background models owing to the limited number of pixels that were downlinked from the spacecraft.  Detection and confirmation of small planet candidates on year-long orbits, and thus estimates of the occurrence rates, would particularly benefit from additional research into the removal of this noise component, or its probabilistic modeling.
\\
\begin{center}
\commentlink{10}
\end{center}
\ \\

\subsubsection{Occurrence rates based on probabilistic catalogs} 
Near the Kepler detection limit, where small planets in long-period orbits are found, detection reliability drops significantly and the confident identification of planet candidates becomes problematic.  Distinguishing both instrumental and astrophysical false positives from true planets becomes very difficult: at low Signal-To-Noise (SNR) many techniques that identify astrophysical false positives struggle, and instrumental systematics can closely mimic small-planet transit signals.  Methods that exclude such false positives often also tend to exclude planets.

Research described in \ref{sec:occRate}, \ref{sec:fwModelPlanets}, and \ref{sec:oneYearOrbits} will alleviate some, but not all, of these issues. The development of vetting techniques that assign planet candidate probabilities to all detections when the SNR is low is a promising, though challenging, approach that would extend \citet{morton2016} to all detections.  Population-based inference techniques using such a probabilistic catalog, extending, for example, \cite{farr2015}, would allow inferences even when all detections have a low probablility of being a planet.  This approach would enable a higher-confidence estimate of the occurrence of terrestrial planets in the habitable zone than is currently available.
\\
\begin{center}
\commentlink{11}
\end{center}
\ \\

\subsection{Stellar astrophysics projects}

\subsubsection{Building a homogeneous catalog of eclipsing binaries and oscillating stars}

Eclipsing binaries can be used to determine precise stellar radii and masses for the purpose of benchmarking stellar models. Oscillating stars can be used to determine precise ages for the purpose of better understanding exoplanet properties and studying the history of the galaxy. Catalogs of these objects provide easy access to users and the opportunity for ensemble analyses of a large number of targets.

The Kepler Eclipsing Binary Working Group have provided an extensive catalog of eclipsing binaries discovered in Kepler data \citep[e.g.][]{ebs1, ebs2}. Cataloging the eclipsing binaries in K2 has also been attempted for early campaigns \citep[e.g.][]{lacourse2015,barros2016,maxted2018,bayliss2018}. So far the K2 archive has not been searched in a complete and homogeneous way -- to our knowledge -- for objects of interest to stellar physics. With the combination of new, advanced machine learning techniques \citep[e.g. Self-Organizing Maps;][]{armstrong2016}, the final (re-processed) data releases, and human-labeled training sets from projects such as Planet Hunters, producing catalogs of variable stars is a tractable and valuable project.
\\
\begin{center}
\commentlink{12}
\end{center}
\ \\

\subsubsection{Comparing Kepler's 29 star clusters}
Star clusters provide the opportunity to study stars that can be assumed to have the same age, composition, and formation history. We can use them as laboratories for understanding stellar evolution and planet formation under controlled conditions. 
Kepler and K2 have observed two star forming associations ($<$10 Myr old), 17 open clusters (1 Myr to 8 Gyr), and nine globular clusters ($\gtrapprox$11 Gyr). Of these, only 7 have been analyzed by 5 or more scientific publications to date \citep[see][for an overview]{cody2018b}. Notable clusters which appear to have been under-utilized are the young Lagoon nebula region (M8), the intermediate-age M35 cluster, and the young Taurus star forming region. 

Several methods exist to analyze the properties of stellar clusters, including asteroseismic analyses \citep[e.g.][]{stello2016}, the identification of eclipsing binary systems to derive benchmark radii and masses \citep[e.g.][]{gillen2017,kraus2017,sandquist2018,torres2018}, and rotation rate studies as a function of age and mass.

Recent analyses of young clusters by K2 have already revealed that late M-type dwarf stars shed angular momentum after star formation in a way that is significantly slower than their earlier-type counterparts \citep[e.g.][]{douglas2017,rebull2018}.
Magnetic fields have recently been suggested to explain these findings \citep{garraffo2018}.
\\
\begin{center}
\commentlink{13}
\end{center}
\ \\

\subsubsection{Capitalizing on contemporaneous color photometry}
Contemporaneous color photometry can unlock the potential of studies of stellar variability and dusty disks, as well as enabling the identification of exoplanet false positives. During K2 Campaigns 16 and 17, Kepler's observations were complemented by contemporaneous PanSTARRS1 photometry from the ground \citep{dotson2018}. PanSTARRS surveyed Kepler's entire field of view for 56 nights in four filters (g, r, i, z).
The primary motivation behind obtaining these data was to identify supernovae in the field in time for contemporaneous follow-up observations, but the high-cadence color data are expected to enable a range of additional studies. The data were made public in September 2018.  High-cadence contemporaneous color photometry was also obtained during K2 Campaign 9 \citep{henderson2016,zang2018}. To date, these data have been under-utilized in conjunction with K2 observations.
\\
\begin{center}
\commentlink{14}
\end{center}
\ \\

\subsubsection{Using rotation rates to map recent star birth}

The rotation rate of stars can inform models of stellar age and identify young stars that have recently formed. Using rotation periods of the 16,000 stars in the Cygnus field observed with Kepler and Gaia parallaxes, \citet{davenport2018} recently reported an excess of fast-rotating, young stars at low Galactic height Z, consistent with a recent burst of star formation in the disk. The study only utilized data from Kepler's original mission. It is possible that a similar analysis of all 20 K2 fields may reveal more clues about the recent star formation history in our Galactic neighborhood.  When combined with Gaia data, such analyses may conceivably provide a spatial map of recent star birth, and perhaps reveal our Galaxy's spiral arm density waves.
\begin{center}
\commentlink{15}
\end{center}
\ \\

\subsubsection{Employing asteroseismology to investigate the history of our Galaxy}
Statistical samples of stellar ages and compositions of red giants offer the opportunity to study the structure and evolution of the Galaxy, i.e. they enable {\em Galactic Archaeology} \citep{miglio2013}.
Recent asteroseismic analyses of red giants observed by Kepler have revealed a strong relationship between the ages of the stars from Kepler and their chemical composition as inferred from APOGEE spectra \citep{aguirre2018}. The result was based on original Kepler field alone. It is likely that similar analyses across the 20 fields observed by K2 will reveal new insights into the history of the Milky Way. To date only early data from K2 have been analyzed in this way \citep{stello2017}. Future studies will be aided greatly by the availability of HERMES spectra \citep{wittenmyer2018}, the new Gaia DR2 parallaxes \citep{gaia2018}, and the introduction of machine learning to enhance the data processing \citep{hon2018}.
\\
\begin{center}
\commentlink{16}
\end{center}
\ \\

\subsubsection{Performing asteroseismology in the time domain}
Asteroseismology can unlock stellar properties such as mass and radius independently from other methods. Until recently, asteroseismic analyses were only carried out in the frequency domain, by using Fourier Transforms on time series photometry. Recent research into the use of Gaussian Process models \citep{george,foreman-mackey2017} and Gaussian Process-based Continuous Auto-Regressive Moving Average models \citep[CARMA;][]{farr2018} have been able to reveal asteroseismic information by fitting simple harmonic oscillator models in the time domain. These techniques offer the potential to unlock oscillations at very low signal to noise level, but their use on Kepler data is yet to be explored thoroughly. Validating these methods on Kepler data, where verification of results using the frequency domain is possible, will help establish whether these methods can be used to detect solar-like oscillations in TESS data.\\
\begin{center}
\commentlink{17}
\end{center}
\ \\

\subsection{Extragalactic projects}

\subsubsection{Discovery and analyses of supernovae}
Amongst many science opportunities, supernovae offer the opportunity to investigate stellar composition and understand mechanisms that drive the evolution of the galaxy. However, supernovae are best understood by observing and modeling the initial few hours and days of the event. This is often difficult to obtain, as supernovae are usually only discovered several days after the explosion.

The Kepler and K2 data sets are thought to contain approximately 60 supernovae and supernova-like transients \citep[e.g.][]{narayan2018,snrest2018,snsmith2018}.  In some cases Kepler has captured a full lightcurve starting from before the explosion to many weeks thereafter \citep[e.g.][]{garnavich2016}. Careful analyses of such data may reveal new insights into the progenitors and the early stages of supernova explosions. For example, a recent event detected by K2 was revealed to be an unusual fast-evolving luminous transient \citep{rest2018}.
\\
\begin{center}
\commentlink{18}
\end{center}
\ \\

\subsubsection{Estimating black hole masses in active galaxies}
Measuring the mass of black holes in active galactic nuclei (AGN) can prove challenging. Understanding the mass of black holes in other galaxies informs our models of galaxy formation in the universe. A recent analysis of the Kepler data of an active galaxy suggests that oscillation frequencies in the Kepler lightcurve of an AGN appear predictive of the central black hole mass \citep{smith2018}. The K2 mission has observed dozens of active galaxies, including the very bright BL Lac-type object OJ 287.  These data have been challenging to use because Kepler's instrumental systematics show time-scales which tend to be similar to the oscillation frequencies of AGN \citep{obrien2018}. However, improved detrending methods may enable Kepler's AGN data set to be utilized. For example, PSF photometry may enable a more effective approach towards removing systematics introduced by focus changes and sources drifting out of apertures.
\\
\begin{center}
\commentlink{19}
\end{center}
\ \\

\subsection{Solar System science projects}

\subsubsection{Characterizing small bodies in the Solar System}
Constraining key properties of asteroids, such as rotation rate and shape, can further our understanding of the history of our own solar system.

The K2 mission targeted 366 small Solar System objects including  12 main-belt asteroids, 21 comets, 77 Trans-Neptunian Objects, and 243 Jovian Trojan and Hilda asteroids.
Kepler's exquisite data enable users to detect slow, low-amplitude rotation periods which are challenging to obtain from the ground. Additionally the high cadence and long baseline of K2 data compared to ground based observation greatly increase our ability to accurately measure light curves of these objects for several rotation periods.

Subsets of the data have already been used to study the shape and albedo of Trans-Neptunian Objects \citep{pal2015,pal2016}, characterize the rotational properties of Trojan asteroids \citep{ryan2017,szabo2017}, and study the irregular satellites of Uranus \citep{farkas2017}.
To date, no uniform catalog of small body characteristics based on K2 data has been published. Such study may provide exquisite data on the spin rates and binary fractions of different asteroid populations, potentially revealing new insights into the formation of our Solar System \citep{ryan2017}.
\\
\begin{center}
\commentlink{20}
\end{center}
\ \\

\subsubsection{Characterizing serendipitously observed asteroids}
Beyond the objects that were targeted by K2, many objects moved through the K2 focal plane during observations that were not specifically targeted. This is particularly interesting in superstamps, where large portions of the focal plane, often hundreds of pixels across, were downlinked from the telescope. These large continuous stamps provide the opportunity to measure light curves for serendipitously observed Solar System objects and produce valuable estimates of rotation rates without the need to obtain more ground based data. 

Because K2 observed fields in the ecliptic plane, the data set is thought to contain a significant number of uncatalogued objects. No uniform search for such objects has been carried out across the data archive to date (to our knowledge), although the idea has successfully been demonstrated on the large K2 superstamp masks which were collected to observe Uranus and Neptune \citep{szabo2016,molnar2018}. Discovering new Solar System objects in the K2 data set may yield interesting information on fast moving and Interior-Earth Objects (IEOs), which are difficult to identify from the ground. 
\\
\begin{center}
\commentlink{21}
\end{center}
\ \\

\section{Conclusions}

In the above sections we have presented a non-exhaustive list of important studies which can be executed immediately, using the public data in the Kepler and K2 archives. These projects would provide significant scientific insights into the study of exoplanets, stars, galaxies, and solar system objects, building on the wealth of studies that have already been completed with Kepler data. Figure \ref{fig:publications} shows the growing number of studies using Kepler data, and the projects we have outlined demonstrate just a few of the ways in which the science output of the Kepler mission will continue to grow.

The academic community's ability to extract science from these data has steadily increased over time for a number of reasons. First, all data collected by Kepler is now public. Second, the quality of the existing data sets continues to increase. Third, the community has invented improved data analysis methods and created new software tools which have dramatically enhanced the value of the data \citep[e.g.][]{vanderburg2015, aigrain2016, luger2016}. 

The full Kepler and K2 datasets are now public and hosted at the data archives at MAST and the NASA Exoplanet Archive. These archives host a rich set of data products which enable in-depth analyses that were not previously possible. There are 781,590 unique target masks in the archive to complete these projects and dozens more that have not yet been envisioned. The impacts of these projects will be far reaching across many fields of study in astronomy, from exoplanets to extra-galactic variability, touching upon key aspects of all NASA's three big astrophysics questions; How does the universe work? How did we get here? Are we alone? 

Kepler's science is expected to continue for many years. With these impactful projects still ahead of the community, it is likely that some of Kepler's biggest discoveries are still ahead of us.
\\

\vspace{5mm}

\acknowledgments
\section*{Acknowledgments}
The success of Kepler is the achievement of a large team of exceptional individuals, including many whose hard and essential work has not necessarily been visible or widely recognized. The author list of this draft paper only reflects the names of the individuals who edited this particular manuscript to advertise and applaud Kepler's data set, not those who deserve credit for Kepler data. Any opinions presented in this paper are the personal views of the authors.

This paper is made possible thanks to the vibrant and productive Kepler community who continue to publish exceptional science, collaborate effectively, and enable more open science through releasing public data analysis routines. The authors would like to thank the community for their continued hard work. In particular, we would like to thank the hundreds of Guest Observers who defined K2's science program by writing target proposals, and the dozens of expert referees who carefully reviewed them.

We thank the Mikulski Archive for Space Telescopes (MAST) and the NASA Exoplanet Archive for making Kepler's data products available to the community. MAST/STScI is operated by the Association of Universities for Research in Astronomy, Inc., under NASA contract NAS5-26555. The NASA Exoplanet Archive is operated by the California Institute of Technology, under contract with NASA under the Exoplanet Exploration Program.
Funding for the Kepler and K2 missions is provided by the NASA Science Mission directorate.

Do not go gentle into that good night, Kepler.
\\

\end{document}